\documentclass[letterpaper,twocolumn,american,aps,prb,superscriptaddress]{revtex4}
\usepackage[LGR,T1]{fontenc}
\usepackage[latin9]{inputenc}
\usepackage{amsmath}
\usepackage{amssymb}
\usepackage{graphicx}
\usepackage{color}

\makeatletter
\pdfpageheight\paperheight
\pdfpagewidth\paperwidth

\pdfpageheight\paperheight
\pdfpagewidth\paperwidth

\ProvideTextCommand{\~}{LGR}[1]{\char126#1}

\usepackage{textcomp}

\usepackage{ulem}
\usepackage{lipsum}
\usepackage{babel}

\makeatother

\begin{document}
\title{Relation between equilibrium quantum phase transitions and dynamical
quantum phase transitions in two-band systems}

\author{Yumeng Zeng}
\affiliation{Beijing National Laboratory for Condensed Matter Physics, Institute
of Physics, Chinese Academy of Sciences, Beijing 100190, China}
\affiliation{School of Physical Sciences, University of Chinese Academy of Sciences,
Beijing 100049, China }

\author{Shu Chen}
\email{schen@iphy.ac.cn }
\affiliation{Beijing National Laboratory for Condensed Matter Physics, Institute
of Physics, Chinese Academy of Sciences, Beijing 100190, China}
\affiliation{School of Physical Sciences, University of Chinese Academy of Sciences,
Beijing 100049, China }

\date{\today}
\begin{abstract}

The dynamical quantum phase transition (DQPT) is an important concept in nonequilibrium critical phenomena; however, its relation to the equilibrium quantum phase transition (EQPT) remains obscure. Substantial evidence has suggested that quenching across the underlying equilibrium phase boundary is neither a sufficient nor a necessary condition for the existence of the DQPT. In this work, we give a necessary and sufficient condition for the occurrence of DQPTs in two-band systems by introducing the quench fidelity, which is defined as the fidelity between the ground-state wave functions of the prequench and postquench Hamiltonian, and elaborate it by taking the one-dimensional anisotropic XY model as an example. The relation between EQPTs and DQPTs is analyzed in detail in terms of the quench fidelity.

\end{abstract}
\maketitle

\section{Introduction}

The equilibrium quantum phase transition (EQPT) is caused by the quantum fluctuations and controlled by the parameters of the Hamiltonian of the system \cite{Sachdev}. The existence of the EQPT reveals that quantum systems with parameters in different equilibrium quantum phases have distinct ground-state properties \cite{Zanardi2006,You2007}. The fidelity, which is the module of the overlap of two ground states of Hamiltonians with different parameter values, is widely used to describe the nonanalytic changes of the properties of ground states around EQPT points \cite{Zanardi2006,You2007,Zanardi2007,Buonsante2007,Cozzini2007-1,Cozzini2007-2,YanfMF2007,ZhouHQ2008,ChenS2008,ZRams2011,Zeng2024,Dutta,SunG2015,SunG}. Using $\gamma=(\gamma_{1},\gamma_{2},\ldots)$ to denote phase-driving parameters of the quantum system, the fidelity can be formulated as
\begin{equation}
\mathcal{F}(\gamma,\tilde{\gamma})=\left|\left\langle \psi^{0}(\gamma)\right|\left.\psi^{0}(\tilde{\gamma})\right\rangle \right|,\label{eq:Fk}
\end{equation}
where $\left|\psi^{0}(\gamma)\right\rangle $ denotes the ground state of the Hamiltonian $H(\gamma)$ and $\tilde{\gamma}$ represents parameters with values different from $\gamma$. A previous work has unveiled that $\mathcal{F}(\gamma,\tilde{\gamma})$ has (does not have) exact zeros in the thermodynamical limit for $\gamma$ and $\tilde{\gamma}$ in different (the same) phases \cite{Zeng2024}. To eliminate the effect of the Anderson orthogonality catastrophe \cite{Anderson1967} in the thermodynamical limit, it is natural to introduce the decay rate function given by
\begin{equation}
\alpha(\gamma,\tilde{\gamma})=-\frac{1}{L}\ln\mathcal{F}(\gamma,\tilde{\gamma}),
\end{equation}
where $L$ is the size of the system. Fixing the value of $\gamma$ and changing the value of $\tilde{\gamma}$, $\alpha(\tilde{\gamma})$ will exhibit nonanalytical behavior in the thermodynamical limit at EQPT points. Thus the decay rate function $\alpha(\tilde{\gamma})$ is a valid tool to probe the EQPT.

Since the dynamical quantum phase transition (DQPT) was proposed more than a decade ago \cite{Heyl2013PRL}, it has been widely recognized that the
DQPT is closely related to the EQPT \cite{Karrasch2013PRB,Kriel2014PRB,Heyl2015RPL,Dora2015PRB,Schmitt2015PRB,HuangZ2016PRB,Zvyagin2016LTP,Bhattacharya2017PRB,Bhattacharjee2018PRB,Heyl2018RPP,Cheraghi2023NJP,Corps,ZhouBZ2019}. The DQPT is a phenomenon in nonequilibrium systems which indicates that the time-evolution state evolves into an orthogonal state of the initial state. As an analog of the fidelity in EQPT theory, the Loschmidt echo (LE) is a key concept in DQPT theory. For an initial state $|\psi_{i}\rangle$ which is the ground state of the initial Hamiltonian $H_{i}$ and its time evolution state $|\psi(t)\rangle=e^{-iH_{f}t}|\psi_{i}\rangle$ which is driven by a postquench Hamiltonian $H_{f}$, the LE is defined as
\begin{equation}
\mathcal{L}(t)=|\langle\psi_{i}|\psi(t)\rangle|^{2},
\end{equation}
which represents the return probability of the time-evolution state to the initial state. Similar to the fidelity across EQPT points, the LE has exact zeros at critical times when DQPTs happen \cite{ZhouBZ2021,ZengYM}. Similarly, to remove the effect of system size properly, the rate function of the LE is introduced as
\begin{equation}
\lambda(t)=-\frac{1}{L}\ln\mathcal{L}(t),\label{rf}
\end{equation}
where $L$ is the size of the system. Evidently, the exact zero of the LE $\mathcal{L}(t)$ can give rise to the singularity of the rate function $\lambda(t)$. Therefore, DQPTs can be characterized by nonanalytic behaviors of the rate function $\lambda(t)$ of the LE in the real-time evolution.

In early works \cite{Heyl2013PRL,Karrasch2013PRB,Kriel2014PRB,Bhattacharjee2018PRB}, it was commonly believed that the condition for the occurrence of DQPTs is consistent with EQPTs; i.e., if the phase-driving parameters are suddenly quenched across the underlying equilibrium phase boundary at the initial time $t=0$, DQPTs will arise at certain critical times. In noninteracting topological systems, it is clear that DQPTs have to occur if the topological number changes under the quench in 1D systems \cite{Dora2015PRB,YangC}, while the occurrence of DQPTs is related to the change in the absolute value of the Chern number in 2D systems \cite{HuangZ2016PRB}. These types of DQPTs are robust and are therefore termed topologically protected. However, it has been discovered that certain particular models do have DQPTs for quenches within the same equilibrium
phase, and even for very weak quenches, which are termed \textit{accidental} DQPTs \cite{Dora2015PRB,Liska,Flaschner,Heyl2018RPP,WongCY2024PRB} and  \textit{anomalous} DQPTs \cite{Homrighausen2017,Halimeh2020}. In fact, quenching across the underlying equilibrium phase boundary is neither a sufficient nor a necessary condition for the existence of the DQPT \cite{Vajna2014PRB,Andraschko2014PRB,Sharma2015PRB,Divakaran2016PRE,Obuchi2017PRB,Halimeh2017PRB,Zauner-Stauber2017PRE,Jafari2019SR,Sadrzadeh2021PRB,Stumper2022PRR,Rossi2022PRB}. This means that a DQPT may not happen even quenching across the underlying EQPT point. The intrinsic relation between EQPTs and DQPTs is an intriguing question. Although some works \cite{Lang2018PRB,Uhrich2020PRB} have discussed this issue, a theoretical study to uncover clearly their relation is necessary.

In this paper, we expose the relation between EQPTs and DQPTs in a general two-band system from a new perspective by introducing a concept of the quench fidelity. In terms of the quench fidelity,  we give a universal necessary and sufficient condition for the occurrence of DQPTs. Then, we demonstrate our theory by taking the XY model as a concrete example and calculating it in detail. Through analysis of the distribution of values of each $k$-mode of the quench fidelity, we also propose a sufficient condition for the occurrence of DQPTs. Finally, we discuss the relationship between quenching across the equilibrium phase transition point and the existence of DQPTs.

\section{The quench fidelity}

In order to discuss DQPTs and EQPTs in the same framework, we will introduce the definition of the quench fidelity and reveal its role in connecting EQPTs and DQPTs.

To start with, we consider a general two-band system of which the Hamiltonian in momentum space can be expressed as
\begin{equation}
H_{k}(\gamma)=\mathbf{d}_{k}(\gamma)\cdot\boldsymbol{\mathbf{\sigma}}+d_{0,k}(\gamma)\mathbb{I},\label{eq:Hk}
\end{equation}
where $H_{k}(\gamma)$ is the Hamiltonian of $k$-mode with momentum $k$, $\boldsymbol{\mathbf{\sigma}}=(\sigma_{x},\sigma_{y},\sigma_{z})$ are Pauli matrices, $\mathbf{d}_{k}(\gamma)=(d_{x,k}(\gamma),d_{y,k}(\gamma),d_{z,k}(\gamma))$ and $d_{0,k}(\gamma)$ are corresponding vector components of $H_{k}(\gamma)$, and $\mathbb{I}$ denotes the unit matrix. We choose $H_{i}=H(\gamma_{i})$ as the initial Hamiltonian, where $\gamma_{i}$ is the prequench parameter. At time $t=0$, we suddenly quench the value of $\gamma$ from $\gamma_{i}$ to $\gamma_{f}$, i.e., $\gamma=\gamma_{i}\Theta(-t)+\gamma_{f}\Theta(t)$. Then the LE of the system which is evolved under the postquench Hamiltonian $H_{f}=H(\gamma_{f})$ can be represented as
\begin{equation}
\mathcal{L}(t)=\prod_{k}\mathcal{L}_{k}(t)=\prod_{k}\left|\langle\psi_{i,k}^{0}|e^{-iH_{f,k}t}|\psi_{i,k}^{0}\rangle\right|^{2},
\end{equation}
where $H_{f,k}$ is the Hamiltonian of the $k$-mode with parameter $\gamma_{f}$, and $|\psi_{i,k}^{0}\rangle$ is the ground state of the prequench Hamiltonian $H_{i,k}$ of the $k$-mode with parameter $\gamma_{i}$. Then we have the $k$-mode of the LE:
\begin{equation}
\mathcal{L}_{k}(t)=1-[1-(\hat{\mathbf{d}}_{k}(\gamma_{i})\cdot\mathbf{\hat{d}}_{k}(\gamma_{f}))^{2}]\sin^{2}(\left|\mathbf{d}_{k}(\gamma_{f})\right|t),\label{eq:Lk}
\end{equation}
where $\mathbf{\hat{d}}_{k}(\gamma_{i/f})=\mathbf{d}_{k}(\gamma_{i/f})/\left|\mathbf{d}_{k}(\gamma_{i/f})\right|$ and $\left|\mathbf{d}_{k}(\gamma_{i/f})\right|=\sqrt{d^{2}_{x,k}(\gamma_{i/f})+d^{2}_{y,k}(\gamma_{i/f})+d^{2}_{z,k}(\gamma_{i/f})}$.

The occurrence of the DQPT requires at least one $k_{c}$-mode of the LE equal to 0, i.e.,  $\mathcal{L}_{k_{c}}(t_{c,n})=0$, where $n$ is a positive integer and $t_{c,n}=(2n-1)\pi/(2\left|\mathbf{d}_{k_{c}}(\gamma_{f})\right|)$ are critical times when DQPTs happen \cite{ZengYM}. Since $\sin^{2}(\left|\mathbf{d}_{k_{c}}(\gamma_{f})\right|t_{c,n})$ always equals 1, it is easy to get $\mathcal{L}_{k_{c}}(t_{c,n})=(\hat{\mathbf{d}}_{k_{c}}(\gamma_{i})\cdot\mathbf{\hat{d}}_{k_{c}}(\gamma_{f}))^{2}$. Evidently, $\mathcal{L}_{k_{c}}(t_{c,n})=0$ is fulfilled if and only if $\hat{\mathbf{d}}_{k_{c}}(\gamma_{i})\cdot\mathbf{\hat{d}}_{k_{c}}(\gamma_{f})=0$; i.e., $\hat{\mathbf{d}}_{k_{c}}(\gamma_{i})$ and $\mathbf{\hat{d}}_{k_{c}}(\gamma_{f})$ are perpendicular on the Bloch sphere \cite{Dora2015PRB}. Whereas what we care about is whether such a $k_{c}$ exists for various values of $\gamma_{i}$ and $\gamma_{f}$, we take no account of values of critical times $t_{c,n}$ when $\mathcal{L}_{k_{c}}$ equals zero, and investigate properties of $\bar{\mathcal{L}}_{k}=\mathcal{L}_{k}(\pi/[2\left|\mathbf{d}_{k}(\gamma_{f})\right|])$ which is the minimum of $\mathcal{L}_{k}(t)$ in time instead of $\mathcal{L}_{k}(t)$. According to Eq. (\ref{eq:Lk}), it is easy to get
\begin{equation}
\bar{\mathcal{L}}_{k}=(\hat{\mathbf{d}}_{k}(\gamma_{i})\cdot\mathbf{\hat{d}}_{k}(\gamma_{f}))^{2},\label{eq:LkT}
\end{equation}
which is time independent and equals zero only for the $k_{c}$-mode, i.e., $\bar{\mathcal{L}}_{k_{c}}=0$.

Now, we introduce the definition of the quench fidelity:
\begin{equation}
\mathcal{F}^{\text{q}}=\prod_{k}\mathcal{F}^{\text{q}}_{k}=\prod_{k}\left|\left\langle \psi_{i,k}^{0}\right|\left.\psi_{f,k}^{0}\right\rangle \right|, \label{eq:Fqk}
\end{equation}
where $|\psi_{i,k}^{0}\rangle$ and $|\psi_{f,k}^{0}\rangle$ are ground states of prequench Hamiltonian $H_{i,k}$ and postquench Hamiltonian $H_{f,k}$,
respectively. Comparing Eq. (\ref{eq:Fqk}) with Eq. (\ref{eq:Fk}), the difference is that the two parameters of the quench fidelity are fixed to the prequench parameter and postquench parameter, respectively. Then we have the $k$-mode of the quench fidelity:
\begin{equation}
\mathcal{F}^{\text{q}}_{k}=\sqrt{\frac{1}{2}(1+\hat{\mathbf{d}}_{k}(\gamma_{i})\cdot\mathbf{\hat{d}}_{k}(\gamma_{f}))}. \label{eq:fidelity}
\end{equation}
For $\gamma_{i}$ and $\gamma_{f}$ being in different phases, $\mathcal{F}^{\text{q}}$ has at least one $k_{0}$-mode that fulfills $\mathcal{F}^{\text{q}}_{k_{0}}=0$; thus there exists at least one $k_{0}$-mode that fulfills $\hat{\mathbf{d}}_{k_{0}}(\gamma_{i})\cdot\mathbf{\hat{d}}_{k_{0}}(\gamma_{f})=-1$; i.e., $\hat{\mathbf{d}}_{k_{0}}(\gamma_{i})$ and $\hat{\mathbf{d}}_{k_{0}}(\gamma_{f})$ are antiparallel on the Bloch sphere. As a consequence, the quench fidelity $\mathcal{F}^{\text{q}}$ equals to zero indicates the occurrence of an EQPT; i.e., $\gamma_{i}$ and $\gamma_{f}$ lie in different phases. If $\gamma_{i}$ and $\gamma_{f}$ are within the same phase, i.e., no EQPT occurs, then $\mathcal{F}^{\text{q}}$ is nonzero.

Comparing Eq. (\ref{eq:fidelity}) with Eq. (\ref{eq:LkT}), it is straightforward to get
\begin{equation}
\bar{\mathcal{L}}_{k}=\left[2(\mathcal{F}^{\text{q}}_{k})^2-1\right]^{2}. \label{eq:relation}
\end{equation}
Equation (\ref{eq:relation}) implies that the $k_{c}$-mode of the LE fulfilling $\bar{\mathcal{L}}_{k_{c}}=0$ corresponds
to the $k_{c}$-mode of $\mathcal{F}^{\text{q}}$ fulfilling $\mathcal{F}^{\text{q}}_{k_{c}}=\sqrt{2}/2$. Therefore, a necessary and sufficient condition for the occurrence of DQPTs is that the quench fidelity of the prequench system and the postquench system has at least a $k_{c}$-mode which satisfies $\mathcal{F}^{\text{q}}_{k_{c}}=\sqrt{2}/2$. For a two-band system, if $d_{0,k_{c}}(\gamma_{f})=0$, then 
\begin{equation}
H_{k_{c}}(\gamma_{f})\left|\psi_{k_{c}}^{0}(\gamma_{i})\right\rangle \propto\left|\psi_{k_{c}}^{1}(\gamma_{i})\right\rangle
\end{equation}
can be derived from $\mathcal{F}_{k_{c}}=\sqrt{2}/2$, where $\left|\psi_{k_{c}}^{1}(\gamma_{i})\right\rangle $ is the first excited state of $H_{k_{c}}(\gamma_{i})$. This implies $\left\langle \psi_{k_{c}}^{0}(\gamma_{i})\right|H_{k_{c}}(\gamma_{f})\left|\psi_{k_{c}}^{0}(\gamma_{i})\right\rangle =0$. And further $\left\langle \psi_{k_{c}}^{0}(\gamma_{i})\right|H_{k_{c}}^{2}(\gamma_{f})\left|\psi_{k_{c}}^{0}(\gamma_{i})\right\rangle =\left|\mathbf{d}_{k_{c}}(\gamma_{f})\right|^{2}$ can be derived. If $d_{0,k_{c}}(\gamma_{i})$ is also equal to 0, then $H_{k_{c}}(\gamma_{i})$ anticommutes with $H_{k_{c}}(\gamma_{f})$, i.e., 
\begin{equation}
\{H_{k_{c}}(\gamma_{i}),H_{k_{c}}(\gamma_{f})\}=0.
\end{equation}
The detailed derivation of these formulas is given in the Appendix \ref{appendix}.


\begin{figure}
\begin{centering}
\includegraphics[scale=0.6]{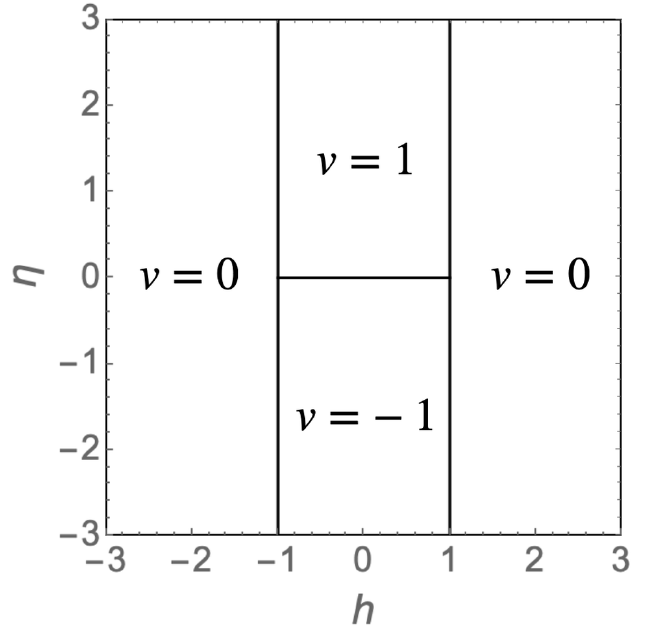}
\par\end{centering}
\caption{The equilibrium phase diagram of the XY model. $\nu$ is the winding number in each phase. \label{Fig-1}}
\end{figure}

\section{ The model and results}

To make our theory more concrete, now we consider exploring the quenching property of a well-known model. Due to the exceptional dynamical phase transition phenomena in the one-dimensional anisotropic XY model \cite{Vajna2014PRB,Liska}, we choose it as an example system to elaborate the relation between EQPTs and DQPTs.

The Hamiltonian of the  one-dimensional spin-$\frac{1}{2}$ anisotropic XY model in a transverse field is described as \cite{Lieb1961AP,Bunder1999PRB,LuoQ2018PRE}
\begin{equation}
H=-\sum_{l=1}^{L}(\frac{1+\eta}{2}\sigma_{l}^{x}\sigma_{l+1}^{x}+\frac{1-\eta}{2}\sigma_{l}^{y}\sigma_{l+1}^{y}+h\sigma_{l}^{z}),
\end{equation}
where $\eta$ is the anisotropy parameter, $h$ denotes the strength of the transverse field, and the periodic boundary condition $\sigma_{L+1}^{x/y}=\sigma_{1}^{x/y}$ is applied. Using the Jordan-Wigner transformation, this model can be mapped to the spinless fermion representation as
\begin{equation}
H=-\sum_{l=1}^{L}(c_{l}^{\dagger}c_{l+1}+\eta c_{l}^{\dagger}c_{l+1}^{\dagger}+\text{H.c.})-\sum_{l=1}^{L}h(2c_{l}^{\dagger}c_{l}-1).
\end{equation}
Here we adopt the periodic boundary condition $c_{L+1}=c_{1}$ and take even $L$. After  the Fourier transformation $c_{l}^{\dagger}=\frac{1}{\sqrt{L}}\sum_{k}e^{ikl}c_{k}^{\dagger}$, the Hamiltonian in the momentum space can be written as
\begin{equation}
H=\sum_{k\in\mathcal{K}}\Psi_{k}^{\dagger}H_{k}\Psi_{k}-2(h+1)c_{0}^{\dagger}c_{0}-2(h-1)c_{\pi}^{\dagger}c_{\pi}+2h,
\end{equation}
where $\Psi_{k}=(c_{k},c_{-k}^{\dagger})^{T}$, $\mathcal{K}=\left\{\frac{2\pi m}{L}\mid m=1,2,\dots,\frac{L-2}{2}\right\}$, and $H_{k}$ takes the form of Eq. (\ref{eq:Hk}) with the vector components expressed as
\begin{align}
d_{z,k} & =-2h-2\cos k, \label{dz}\\
d_{y,k} & =-2\eta\sin k, \label{dy}\\
d_{x,k} &= d_{0,k}=0. \label{dx}
\end{align}
The ground state phase diagram of the XY model is shown in Fig. \ref{Fig-1}. There are three equilibrium phase boundaries: the critical line $h=-1$ where the energy gap closes at $k=0$, the critical line $h=1$ where the energy gap closes at $k=\pi$, and the critical line $\eta=0$ for $-1<h<1$ where the energy gap closes at $k=\pm\arccos(-h)$. The winding number $\nu$ is defined by $\nu=\frac{1}{2\pi} \int_{-\pi}^{\pi} dk \frac{d_{x,k} \partial_k d_{y,k} - d_{y,k} \partial_k d_{x,k} }{d_{x,k}^2+d_{y,k}^2}$.

Choose $\gamma=(h,\eta)$ as quench parameters, and quench them from $\gamma_{i}=(h_{i}, \eta_{i})$ to $\gamma_{f}=(h_{f}, \eta_{f})$ at time $t=0$. Substituting the expression of $\mathbf{d}_{k}(\gamma_{i/f})$ given by Eqs. (\ref{dz})-(\ref{dx}) into Eq. (\ref{eq:fidelity}) and Eq. (\ref{eq:relation}),  we can get the expressions of $\mathcal{F}_{k}^{\text{q}}$ and $\bar{\mathcal{L}}_{k}$, which are explicitly given by
\begin{widetext}
\begin{align}
\mathcal{F}_{k}^{\text{q}}=&\sqrt{\frac{(h_{i}+\cos k)(h_{f}+\cos k)+\eta_{i}\eta_{f}\sin^{2}k}{2\sqrt{[(h_{i}+\cos k)^{2}+\eta_{i}^{2}\sin^{2}k][(h_{f}+\cos k)^{2}+\eta_{f}^{2}\sin^{2}k]}}+\frac{1}{2}},\label{eq:xyFk}
\\
\bar{\mathcal{L}}_{k}=&\frac{[(h_{i}+\cos k)(h_{f}+\cos k)+\eta_{i}\eta_{f}\sin^{2}k]^{2}}{[(h_{i}+\cos k)^{2}+\eta_{i}^{2}\sin^{2}k][(h_{f}+\cos k)^{2}+\eta_{f}^{2}\sin^{2}k]}. \label{eq:xyLk}
\end{align}
\end{widetext}

\begin{figure}
\begin{centering}
\includegraphics[scale=0.52]{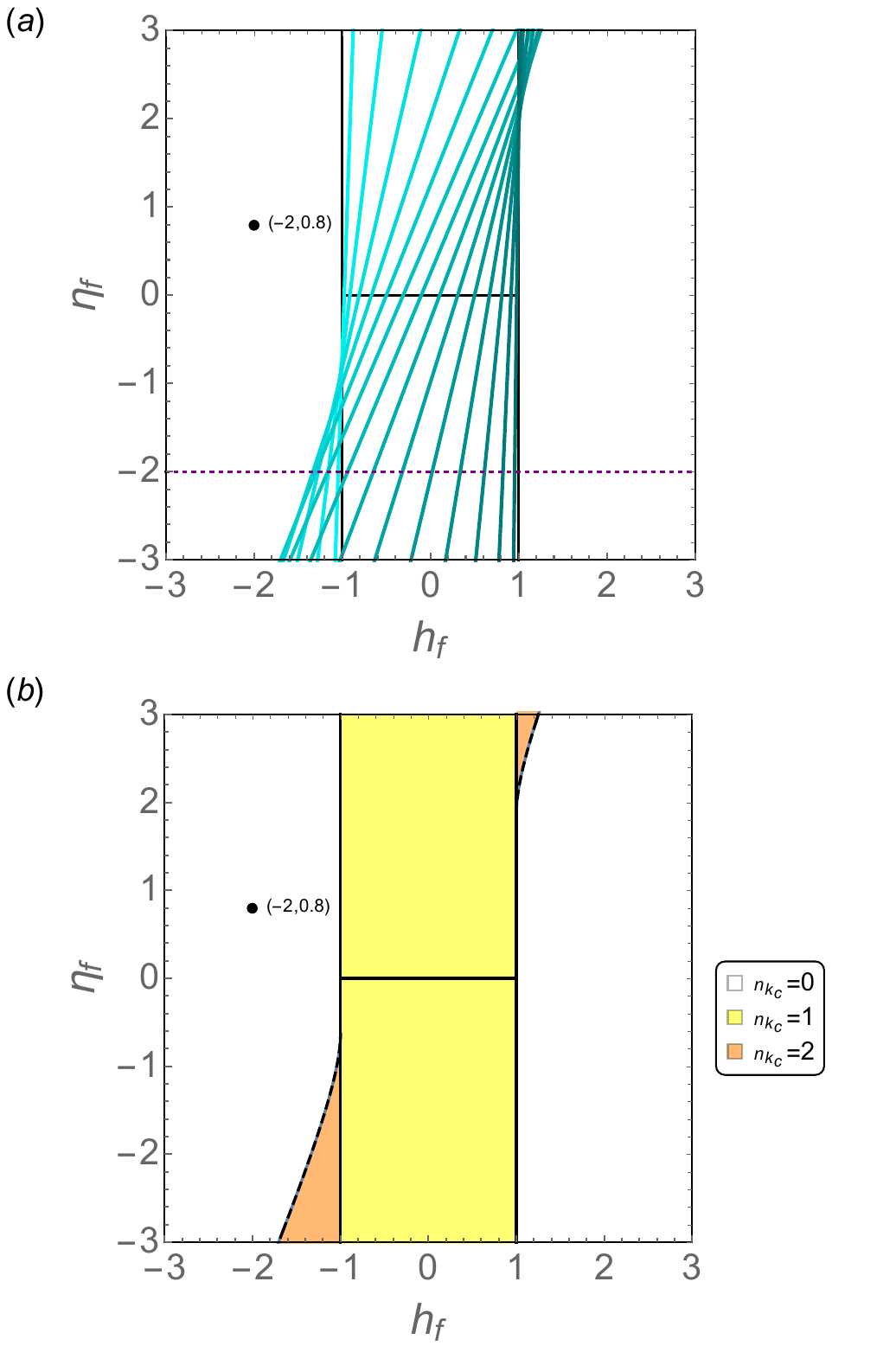}
\par\end{centering}
\caption{(a) The distribution of $k_{c}$-modes $\bar{\mathcal{L}}_{k_{c}}$ in the postquench parameter space spanned by $h_{f}$ and $\eta_{f}$. The cyan solid lines are postquench parameters which possess $\bar{\mathcal{L}}_{k}=0$ with $k=2\pi/L, 4\pi/L, \dots, (L-2)\pi/L$ from lighter to darker, respectively. The purple dotted line denotes $\eta_{f}=-2$. The black dot denotes point $(-2,0.8)$. Here $L=30$. (b) The dynamical phase diagram of DQPTs of the XY model with different numbers $n_{k_{c}}$ of the LE. DQPTs will occur when postquench parameters are in yellow and orange regions. Here $h_{i}=-2$ and $\eta_{i}=0.8$. \label{Fig-2}}
\end{figure}

To clearly exhibit the distribution of postquench parameters which possess $\bar{\mathcal{L}}_{k}=0$ for $h_{i}=-2$ and $\eta_{i}=0.8$, we plot Fig. \ref{Fig-2}(a) according to Eq. (\ref{eq:xyLk}) for the case $L=30$, where the cyan solid lines are postquench parameters which possess $\bar{\mathcal{L}}_{k}=0$ with $k=2\pi/L, 4\pi/L, \dots, (L-2)\pi/L$ from lighter to darker, respectively.  Each postquench parameter point has as many $k_{c}$-modes which satisfy $\bar{\mathcal{L}}_{k_c}=0$ as the number of cyan solid lines that pass through it. As the system size increases, the cyan solid lines become denser and denser. According to the number $n_{k_{c}}$ of $k_{c}$-modes owned by the postquench parameter points, we obtain the dynamical phase diagram of DQPTs of the XY model in the thermodynamic limit for $h_{i}=-2$ and $\eta_{i}=0.8$ as shown in Fig. \ref{Fig-2}(b). Compared with Fig. \ref{Fig-1}, the two additional dashed lines in Fig. \ref{Fig-2}(b) demonstrate differences between dynamical phase boundaries and equilibrium phase boundaries. Obviously, DQPTs will occur when postquench parameters are in yellow and orange regions in Fig. \ref{Fig-2}(b). Since $k_{c}$-modes may exist for postquench parameters and prequench parameters being in the same phase, even quenching within the same equilibrium phase may also lead to the occurrence of DQPTs.

\begin{figure}
\begin{centering}
\includegraphics[scale=0.56]{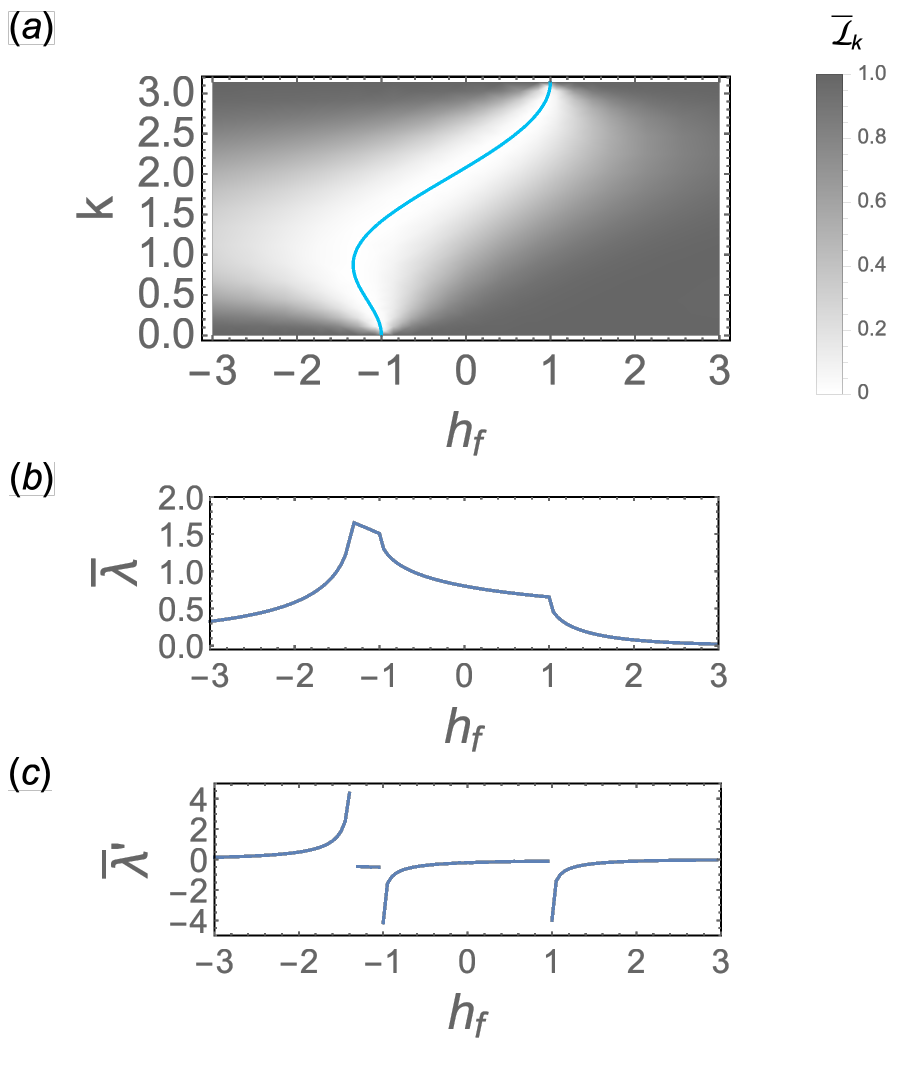}
\par\end{centering}
\caption{ (a) The density graphic of $\bar{\mathcal{L}}_{k}$ versus $k$ and $h_{f}$. The cyan solid line represents $\bar{\mathcal{L}}_{k}=0$. (b) The image of the rate function $\bar{\lambda}$ versus $h_{f}$. (c) The image of the derivative of the rate function $\bar{\lambda}'$ versus $h_{f}$. Here $h_{i}=-2$, $\eta_{i}=0.8$,  and $\eta_{f}=-2$. \label{Fig-3}}
\end{figure}

Setting $\eta_{f}=-2$, we plot the density graphic of $\bar{\mathcal{L}}_{k}$ versus $k$ and $h_{f}$ in Fig. \ref{Fig-3}(a). The cyan solid line represents $\bar{\mathcal{L}}_{k}=0$. As seen from Fig. \ref{Fig-3}(a), whenever postquench parameters cross dynamical phase boundaries of DQPTs as depicted in Fig. \ref{Fig-2}(b), there is a change in the number of $k_{c}$-modes but no abrupt change of $\bar{\mathcal{L}}_{k}$. Figures \ref{Fig-3}(b) and \ref{Fig-3}(c) exhibit images of the rate function $\bar{\lambda}=-\frac{1}{L}\sum_{k}\ln\bar{\mathcal{L}}_{k}$ and its derivative $\bar{\lambda}'$ with respect to $h_{f}$. The rate function $\bar{\lambda}$ exhibits nonanalyticity at all three dynamical phase boundaries. The derivative $\bar{\lambda}'$ becomes divergent when approaching a dynamical phase boundary from a region with fewer $k_{c}$-modes to a region with more $k_{c}$-modes, while it remains finite at the other side. We cannot distinguish the additional phase boundary of DQPTs from the two boundaries that coincide with equilibrium phase boundaries through the image of the rate function, because the abrupt changes of $k$-modes $\mathcal{F}_{k}^{\text{q}}$ with $k=0$ and $k=\pi$ from 1 to 0 or from 0 to 1 at the two equilibrium phase boundaries are unchanged for $\bar{\mathcal{L}}_{k}$ according to Eq. (\ref{eq:relation}). There is only a change of the number of $k_{c}$-modes but no abrupt change of $\bar{\mathcal{L}}_{k}$ at all three boundaries.

\begin{figure}
\begin{centering}
\includegraphics[scale=0.49]{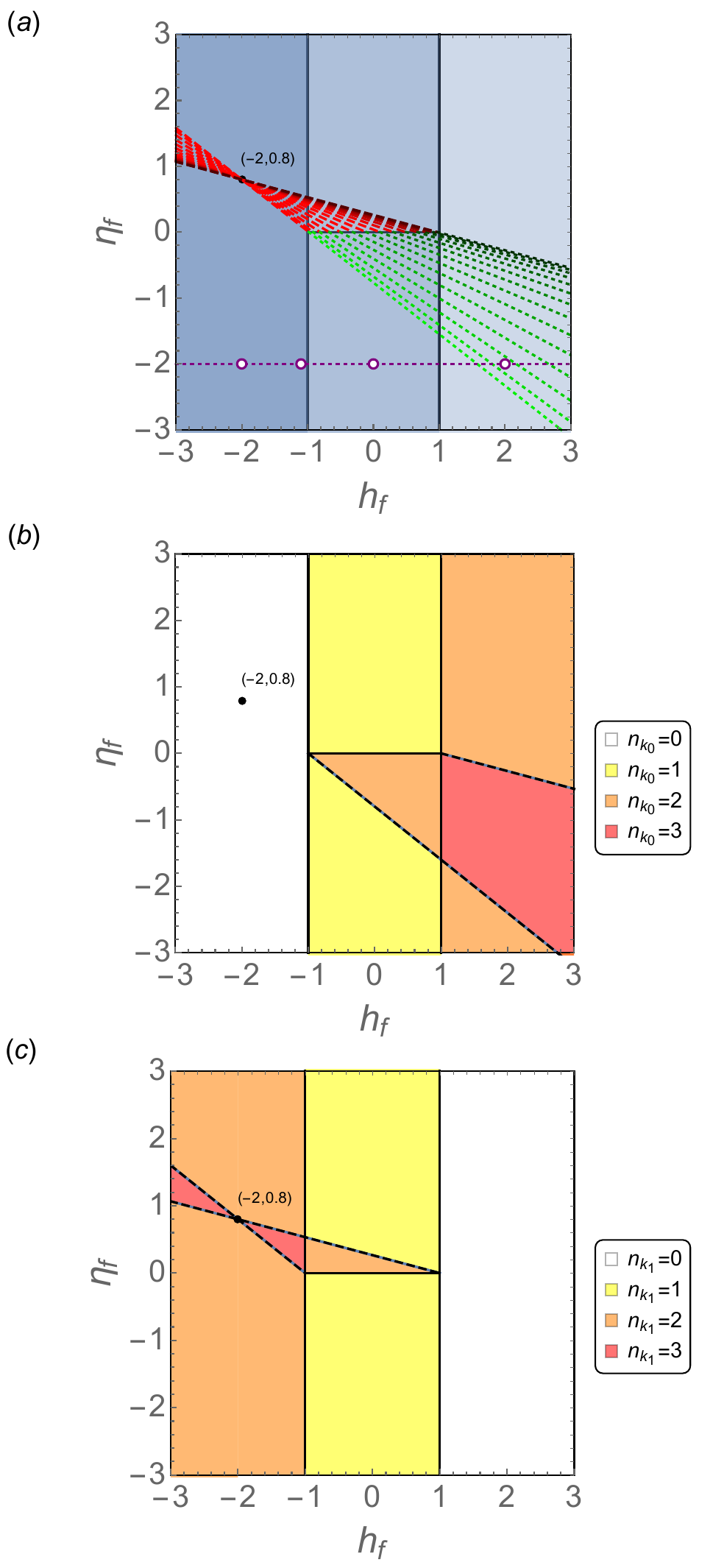}
\par\end{centering}
\caption{(a) The distribution of $k_{0}$-modes $\mathcal{F}_{k_{0}}^{\text{q}}$ and $k_{1}$-modes $\mathcal{F}_{k_{1}}^{\text{q}}$ in the postquench parameter space spanned by $h_{f}$ and $\eta_{f}$. The red dashed lines and green dotted lines depict postquench parameters which possess $\mathcal{F}_{k}^{\text{q}}=1$ and $\mathcal{F}_{k}^{\text{q}}=0$ with $k=2\pi/L, 4\pi/L, \dots, (L-2)\pi/L$ from lighter to darker, respectively. The postquench parameters in dark blue region $h_{f}<-1$ possess both $\mathcal{F}_{0}^{\text{q}}=1$ and $\mathcal{F}_{\pi}^{\text{q}}=1$. The postquench parameters in light blue region $-1<h_{f}<1$ possess $\mathcal{F}_{0}^{\text{q}}=0$ and $\mathcal{F}_{\pi}^{\text{q}}=1$. The postquench parameters in faint blue region $h_{f}>1$ possess both $\mathcal{F}_{0}^{\text{q}}=0$ and $\mathcal{F}_{\pi}^{\text{q}}=0$. The purple dotted line denotes $\eta_{f}=-2$. The black dot denotes point $(-2,0.8)$. Here $L=30$. (b) and (c) are regions with different values of $n_{k_{0}}$ and $n_{k_{1}}$, respectively. Solid black lines are original equilibrium phase boundaries, and dashed black lines are additional boundaries. Here $h_{i}=-2$ and $\eta_{i}=0.8$. \label{Fig-4}}
\end{figure}

For consistency, we call $k$-modes satisfying $\mathcal{F}_{k}^{\text{q}}=0$ as $k_{0}$-modes and call $k$-modes satisfying $\mathcal{F}_{k}^{\text{q}}=1$ as $k_{1}$-modes in the following. To clearly exhibit how $k$-modes $\mathcal{F}_{k}^{\text{q}}$ abruptly change at equilibrium phase boundaries, we display the distribution of  $k_{0}$-modes $\mathcal{F}_{k_{0}}^{\text{q}}$ and $k_{1}$-modes $\mathcal{F}_{k_{1}}^{\text{q}}$ according to Eq. (\ref{eq:xyFk}) in the space of postquench parameters in Fig. \ref{Fig-4}(a) with system size $L=30$, where fixed prequench parameters $h_{i}=-2$ and $\eta_{i}=0.8$. The red dashed lines and green dotted lines depict postquench parameters which possess $\mathcal{F}_{k}^{\text{q}}=1$ and $\mathcal{F}_{k}^{\text{q}}=0$ with $k=2\pi/L, 4\pi/L, \dots, (L-2)\pi/L$ from lighter to darker, respectively. In addition, for $k=0$, $\mathcal{F}_{0}^{\text{q}}=1$ for $h_{f}<-1$, and $\mathcal{F}_{0}^{\text{q}}=0$ for $h_{f}>-1$. For $k=\pi$, $\mathcal{F}_{\pi}^{\text{q}}=1$ for $h_{f}<1$, and $\mathcal{F}_{\pi}^{\text{q}}=0$ for $h_{f}>1$. It is evident that there are only $k_{1}$-modes but no $k_{0}$-modes for $h_{f}<-1$, there are both $k_{0}$-modes and $k_{1}$-modes for $-1<h_{f}<1$, and there are only $k_{0}$-modes but no $k_{1}$-modes for $h_{f}>1$. From Fig. \ref{Fig-4}(a), we can conclude that each point on equilibrium phase boundaries corresponds to a $k$-mode $\mathcal{F}_{k}^{\text{q}}$ whose value will suddenly change at that point and smoothly change everywhere else. As the system size increases, the red dashed lines and green dotted lines become denser and denser. Setting $h_{i}=-2$ and $\eta_{i}=0.8$,  we divide the postquench parameter space in the thermodynamic limit into four kinds of regions as shown in Figs. \ref{Fig-4}(b) and \ref{Fig-4}(c), according to the number $n_{k_{0}}$ of $k_{0}$-modes which satisfy $\mathcal{F}^{\text{q}}_{k_{0}}=0$ and the number $n_{k_{1}}$ of $k_{1}$-modes which satisfy $\mathcal{F}^{\text{q}}_{k_{1}}=1$, respectively. Although Figs. \ref{Fig-4}(b) and \ref{Fig-4}(c) are a little different from Fig. \ref{Fig-1}, the overlapping boundaries of Figs. \ref{Fig-4}(b) and \ref{Fig-4}(c) match exactly the equilibrium phase boundaries of Fig. \ref{Fig-1}. This implies that both $n_{k_{0}}$ and $n_{k_{1}}$ change at equilibrium phase boundaries. Comparing Fig. \ref{Fig-4}(b) with Fig. \ref{Fig-1}, we find that Fig. \ref{Fig-4}(b) has two extra dashed lines across which there will appear more or fewer $k_{0}$-modes, whereas abrupt change in each $\mathcal{F}_{k}^{\text{q}}$ only occurs at equilibrium phase boundaries and not here, which we will demonstrate later. Besides, positions of these two extra dashed lines depend on the value of $h_{i}$ and $\eta_{i}$ and thus are not robust.

\begin{figure}
\begin{centering}
\includegraphics[scale=0.60]{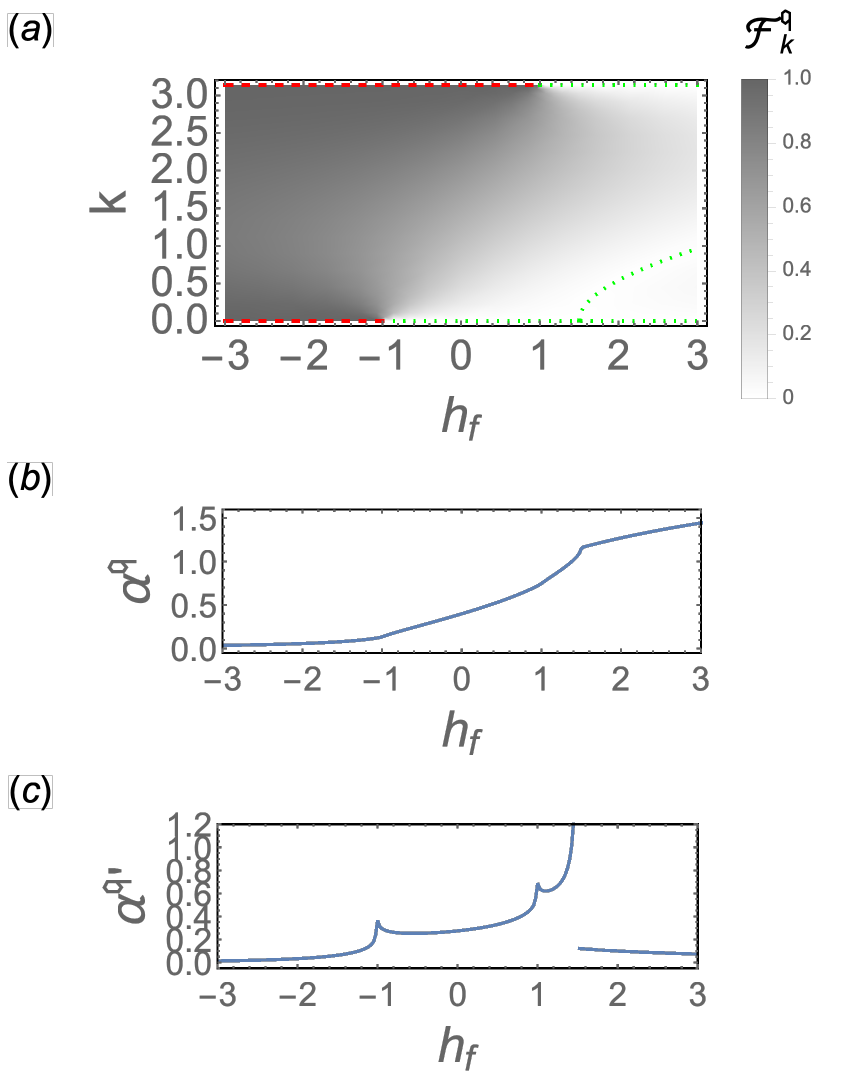}
\par\end{centering}
\caption{(a) The density graphic of $\mathcal{F}_{k}^{\text{q}}$ versus $k$ and $h_{f}$. The red dashed lines and green dotted lines represent $\mathcal{F}_{k}^{\text{q}}=1$ and $\mathcal{F}_{k}^{\text{q}}=0$, respectively. (b) The image of the decay rate function $\alpha^{\text{q}}$ versus $h_{f}$. (c) The image of the derivative of the decay rate function $\alpha^{\text{q}}$$'$ versus $h_{f}$. Here $h_{i}=-2$, $\eta_{i}=0.8$, and $\eta_{f}=-2$.
\label{Fig-5}}
\end{figure}

Setting $\eta_{f}=-2$, we plot the density graphic of $\mathcal{F}_{k}^{\text{q}}$ versus $k$ and $h_{f}$ in Fig. \ref{Fig-5}(a). The red dashed lines represent $\mathcal{F}_{k}^{\text{q}}=1$, and the green dotted lines represent $\mathcal{F}_{k}^{\text{q}}=0$. As seen from Fig. \ref{Fig-5}(a), when the postquench parameter crosses the equilibrium phase boundaries $h_{f}=-1$ and $h_{f}=1$, there is a $k$-mode $\mathcal{F}_{k}^{\text{q}}$ abruptly dropping  from $1$ to $0$. When the parameter crosses the extra boundary in Fig. \ref{Fig-4}(b), there is only a change in the number of $k_{0}$-modes without any abrupt change in  each $\mathcal{F}_{k}^{\text{q}}$. Figures \ref{Fig-5}(b) and \ref{Fig-5}(c) exhibit images of the decay rate function of the quench fidelity $\alpha^{\text{q}}=-\frac{1}{L}\ln\mathcal{F}^{\text{q}}$ and its derivative $\alpha^{\text{q}}$$'$ with respect to $h_{f}$. The decay rate function exhibits nonanalyticity at all three boundaries. Nonetheless, the derivative $\alpha^{\text{q}}$$'$ is divergent around the two equilibrium phase boundaries but is finite at one side of the additional boundary. It is because there is only a change of the number of $k_{0}$-modes but no abrupt change of $\mathcal{F}_{k}^{\text{q}}$ at the extra boundary, while  both of them change at equilibrium phase boundaries. As a result, the behavior of the derivative $\alpha^{\text{q}}$$'$ can distinguish equilibrium phase boundaries from extra boundaries.

\begin{figure}
\begin{centering}
\includegraphics[scale=0.4]{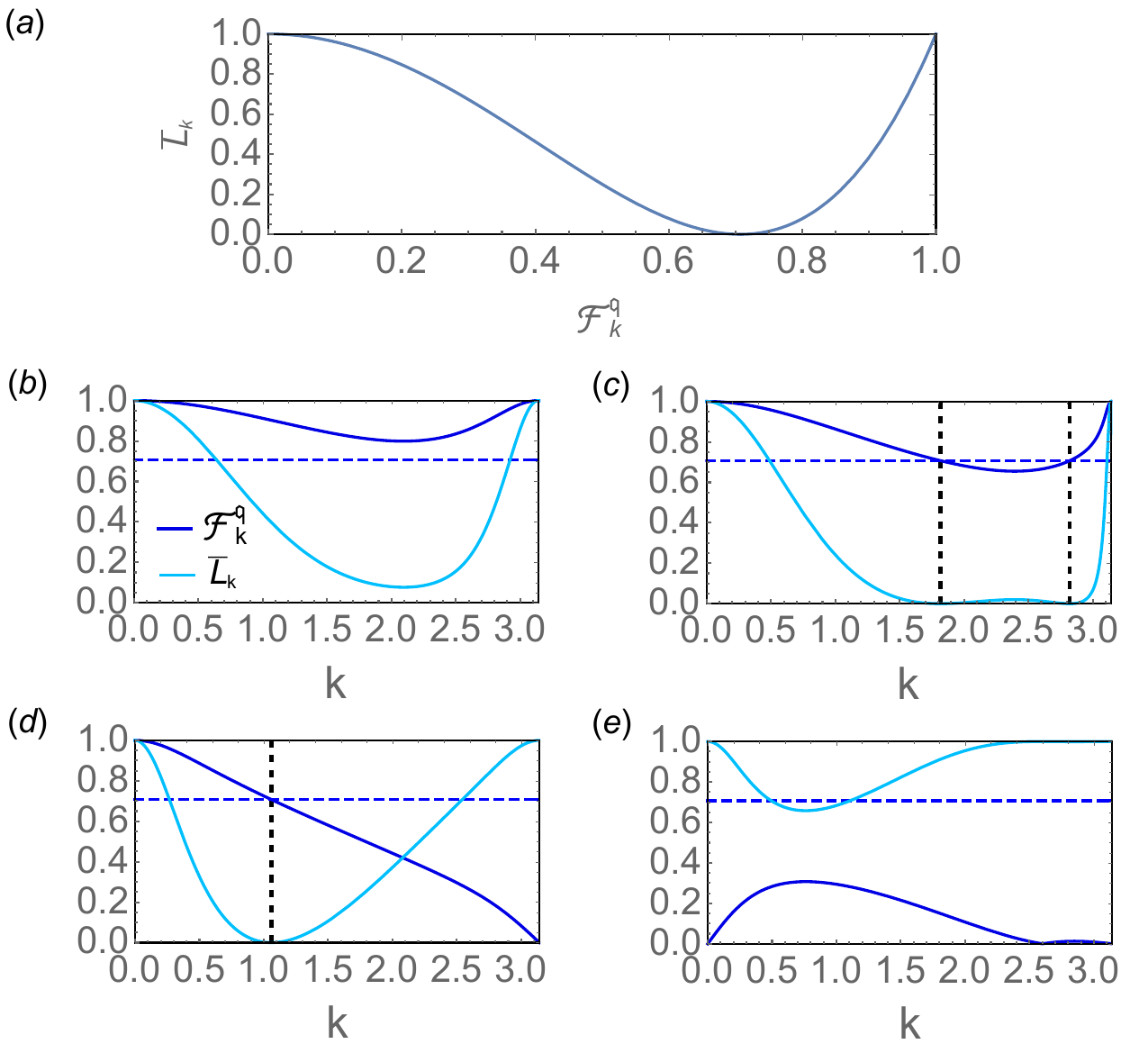}
\par\end{centering}
\caption{(a) The image of $\bar{\mathcal{L}}_{k}$ versus $\mathcal{F}_{k}^{\text{q}}$. (b)--(e) are images of $\mathcal{F}_{k}^{\text{q}}$ and $\bar{\mathcal{L}}_{k}$ versus $k$. The dark blue solid lines are $\mathcal{F}_{k}^{\text{q}}$. The light blue solid lines are $\bar{\mathcal{L}}_{k}$. The blue horizontal dashed lines represent $\sqrt{2}/2$. The black vertical dashed lines denote corresponding $k_{c}$. Postquench parameters are (b) $h_{f}=-2,\ \eta_{f}=-2$; (c) $h_{f}=-1.1,\ \eta_{f}=-2$;
(d) $h_{f}=0,\ \eta_{f}=-2$; (e) $h_{f}=2,\ \eta_{f}=-2$. Here $h_{i}=-2$ and $\eta_{i}=0.8$.  \label{Fig-6}}
\end{figure}

Next we explore the relation between $\mathcal{F}_{k}^{\text{q}}$ and $\bar{\mathcal{L}}_{k}$ of the XY model. According to Eq. (\ref{eq:relation}), we plot the image of $\bar{\mathcal{L}}_{k}$ versus $\mathcal{F}_{k}^{\text{q}}$ in Fig. \ref{Fig-6}(a). It demonstrates that $\bar{\mathcal{L}}_{k}$ must undergo a zero value as long as $\mathcal{F}_{k}^{\text{q}}$ goes from $0$ to $1$. Because $\mathcal{F}_{k}^{\text{q}}$ is a continuous function of $k$, if $\mathcal{F}^{\text{q}}$ has a $k_{0}$-mode satisfying $\mathcal{F}_{k_{0}}^{\text{q}}=0$ and a $k_{1}$-mode satisfying $\mathcal{F}_{k_{1}}^{\text{q}}=1$, then $\mathcal{F}^{\text{q}}$ must have at least a $k_{c}$ -mode between $k_{0}$ and $k_{1}$ which satisfies $\mathcal{F}_{k_{c}}^{\text{q}}=\sqrt{2}/2$  so that $\bar{\mathcal{L}}_{k_{c}}=0$. As a consequence, a sufficient condition for the occurrence of DQPTs is that the quench fidelity $\mathcal{F}^{\text{q}}$ has at least a $k_{0}$-mode and a $k_{1}$-mode simultaneously. To make it more intuitional, we take the four postquench parameter points denoted by purple circles on the purple dashed line in Fig.  \ref{Fig-4}(a) as four cases, and plot images of $\mathcal{F}_{k}^{\text{q}}$ and $\bar{\mathcal{L}}_{k}$ versus $k$ in Figs. \ref{Fig-6}(b)--\ref{Fig-6}(e), respectively. The dark blue and light blue solid lines represent $\mathcal{F}_{k}^{\text{q}}$ and $\bar{\mathcal{L}}_{k}$, respectively. The blue horizontal dashed lines represent $\sqrt{2}/2$, and the black vertical dashed lines denote the corresponding $k_{c}$. Setting $h_{i}=-2$ and $\eta_{i}=0.8$, Figs. \ref{Fig-6}(b) and \ref{Fig-6}(e) show that the quench fidelity only has $k_{1}$-modes and $k_{0}$-modes, respectively, and the images of $\mathcal{F}_{k}^{\text{q}}$ do not cross the horizontal dashed line; thus there is no $k_{c}$-mode satisfying $\bar{\mathcal{L}}_{k_{c}}=0$, which means no DQPT will happen. Figure \ref{Fig-6}(d) shows that the quench fidelity has both a $k_{0}$-mode and a $k_{1}$-mode, and the image of $\mathcal{F}_{k}^{\text{q}}$ certainly crosses the horizontal dashed line; thus $\bar{\mathcal{L}}_{k}$ becomes 0 at the cross point $k_{c}$ and DQPTs can occur. Besides, even if the quench fidelity only has $k_{1}$-modes and has no $k_{0}$-mode, the image of $\mathcal{F}_{k}^{\text{q}}$ can still cross the horizontal dashed line, which leads to the occurrence of DQPTs as shown in Fig. \ref{Fig-6}(c). This is consistent with Fig. \ref{Fig-2} and Fig. \ref{Fig-4}: The quench fidelity has both $k_{0}$-modes and $k_{1}$-modes for $-1<h_{f}<1$, and thus DQPTs must occur when postquench parameters are in this region, while for $h_{f}<-1$ and $h_{f}>1$, the quench fidelity has either only $k_{1}$-modes or only $k_{0}$-modes, and thus DQPTs may or may not occur in these regions. In a nutshell, since a necessary and sufficient condition for DQPTs is that the quench fidelity of the prequench system and the postquench system has at least a $k_{c}$-mode which satisfies $\mathcal{F}_{k_{c}}^{\text{q}}=\sqrt{2}/2$, whereas the necessary and sufficient condition for crossing the EQPT point is the quench fidelity having at least a $k_{0}$-mode which satisfies $\mathcal{F}_{k_{0}}^{\text{q}}=0$, crossing the EQPT point is neither sufficient nor necessary for the occurrence of DQPTs.

\section{Conclusion and discussion}

In summary,  we proposed the concept of the quench fidelity and applied it to unveil the relation between EQPTs and DQPTs from a new perspective. By using $\bar{\mathcal{L}}_{k}$ to denote the minimum of each $\mathcal{L}_{k}(t)$ in time, and defining the quench fidelity $\mathcal{F}^{\text{q}}$ as the module of the overlap of ground states of the Hamiltonian with the prequench parameter and the one with the postquench parameter, we found the relation between $\bar{\mathcal{L}}_{k}$ and $\mathcal{F}_{k}^{\text{q}}$, which reveals that the occurrence of DQPTs corresponds to $k_{c}$-modes of $\mathcal{F}^{\text{q}}$ which satisfy $\mathcal{F}_{k_{c}}^{\text{q}}=\sqrt{2}/2$. Therefore, a necessary and sufficient condition for the existence of DQPTs is that the quench fidelity of the prequench system and the postquench system has at least a $k_{c}$-mode which satisfies $\mathcal{F}_{k_{c}}^{\text{q}}=\sqrt{2}/2$.

To demonstrate our theory, we took the one-dimensional spin-$\frac{1}{2}$ anisotropic XY model as a concrete example and calculated it in detail. According to the number $n_{k_{c}}$ of $k_{c}$-modes, we divided the postquench parameter space into three kinds of regions, and thus got the dynamical phase diagram of DQPTs of the XY model for given prequench parameters. In comparison with the original equilibrium phase diagram, the dynamical phase diagram of the XY model has two additional boundaries, which demonstrates the essential difference between EQPTs and DQPTs.
According to the number $n_{k_{0}}$ of $k_{0}$-modes and the number $n_{k_{1}}$ of $k_{1}$-modes, we divided the postquench parameter space into different regions. Although phase diagrams of $k_{0}$-modes and $k_{1}$-modes are a little different from the original equilibrium phase diagram, the overlapping boundaries of them match exactly the equilibrium phase boundaries, indicating that there are changes in the values of both $n_{k_{0}}$ and $n_{k_{1}}$ at equilibrium phase boundaries. Besides, we found that the phase diagram of $k_{0}$-modes has two extra boundaries across which there will appear more or fewer $k_{0}$-modes, whereas abrupt change in each $\mathcal{F}_{k}^{\text{q}}$ only occurs at equilibrium phase boundaries. The discrepancy between the extra boundaries and equilibrium phase boundaries can be demonstrated by the image of the derivative of the decay rate function of the quench fidelity. Through the image of $\bar{\mathcal{L}}_{k}$ versus $\mathcal{F}_{k}^{\text{q}}$, we concluded that a sufficient condition for the occurrence of DQPTs is that the quench fidelity has at least a $k_{0}$-mode and a $k_{1}$-mode simultaneously, whereas DQPTs may or may not occur if the quench fidelity has only $k_{1}$-modes or only $k_{0}$-modes. This means that quenching across the equilibrium phase point is not sufficient for the occurrence of DQPTs unless the quench fidelity owns at least a $k_{1}$-mode, and quenching within the same equilibrium phase can also give rise to DQPTs, which accounts for the mismatch between dynamical phases and equilibrium phases.

Our work clarifies clearly the relation between EQPTs and DQPTs and provides a new insight for future study on conditions for the occurrence of DQPTs. Nonetheless, our theory is currently limited to two-band systems, and further research on promoting it to a broader range of models is anticipated.

\begin{acknowledgments}
This work is supported by the NSFC under
Grants No. 12174436, No.12474287 and No. T2121001.
\end{acknowledgments}

\appendix
\section{The detailed derivation of formulas}\label{appendix}
For simplicity, here we use $\left|0_{i/f}\right\rangle$ and $\left|1_{i/f}\right\rangle$ to denote the ground state $\left|\psi_{k_{c}}^{0}(\gamma_{i/f})\right\rangle$ and the first excited state $\left|\psi_{k_{c}}^{1}(\gamma_{i/f})\right\rangle$ of $H_{i/f,k_{c}}$, respectively.

For a two-band system, if $d_{0,k_{c}}(\gamma_{f})=0$, then the eigenvalue equations of $H_{f,k_{c}}$ are
\begin{align}
H_{f,k_{c}}\left|0_f\right\rangle & =-E_{f,k_c}\left|0_f\right\rangle,\\
H_{f,k_{c}}\left|1_f\right\rangle & =E_{f,k_c}\left|1_f\right\rangle.
\end{align}
Here $E_{f,k_c}=\left|\mathbf{d}_{k_{c}}(\gamma_{f})\right|$.

Since $\mathcal{F}_{k_{c}}=\left|\left\langle 0_i|0_f\right\rangle\right|=\sqrt{2}/2$, we can derive
\begin{align}
\left|\left\langle 1_i|0_f\right\rangle\right|  & =\sqrt{\left\langle 0_f|1_i\right\rangle\left\langle 1_i|0_f\right\rangle } \nonumber\\
 & =\sqrt{\left\langle 0_f|(1-\left|0_i\right\rangle\left\langle 0_i\right|)|0_f\right\rangle }  \nonumber\\
 & =\sqrt{\left\langle 0_f|0_f\right\rangle-\left\langle 0_f|0_i\right\rangle\left\langle 0_i|0_f\right\rangle }  \nonumber\\
 & =\sqrt{1-\left|\left\langle 0_i|0_f\right\rangle\right|^2 }\nonumber\\
 & =\sqrt{1-1/2 }\nonumber\\
 & = \sqrt{2}/2,
\end{align}
and
\begin{align}
\left|\left\langle 0_i|1_f\right\rangle\right|  & =\sqrt{\left\langle 0_i|1_f\right\rangle\left\langle 1_f|0_i\right\rangle } \nonumber\\
 & =\sqrt{\left\langle 0_i|(1-\left|0_f\right\rangle\left\langle 0_f\right|)|0_i\right\rangle } \nonumber \\
 & =\sqrt{\left\langle 0_i|0_i\right\rangle-\left\langle 0_i|0_f\right\rangle\left\langle 0_f|0_i\right\rangle }  \nonumber\\
 & =\sqrt{1-\left|\left\langle 0_i|0_f\right\rangle\right|^2 }\nonumber\\
 & =\sqrt{1-1/2 }\nonumber\\
 & = \sqrt{2}/2.
\end{align}

Without loss of generality, we assume that the argument of $\left\langle 1_i|0_f\right\rangle\left\langle 0_f|0_i\right\rangle$ is $\theta$, then $\left\langle 1_i|0_f\right\rangle\left\langle 0_f|0_i\right\rangle=e^{i\theta}\left|\left\langle 1_i|0_f\right\rangle\left\langle 0_f|0_i\right\rangle\right|=e^{i\theta}/2$, and thus we can derive
\begin{align}
H_{f,k_{c}}\left|0_i\right\rangle  & =H_{f,k_{c}}(\left|0_f\right\rangle\left\langle 0_f\right|+\left|1_f\right\rangle\left\langle 1_f\right|)\left|0_i\right\rangle \nonumber\\
& =H_{f,k_{c}}\left|0_f\right\rangle\left\langle 0_f|0_i\right\rangle+H_{f,k_{c}}\left|1_f\right\rangle\left\langle 1_f|0_i\right\rangle\nonumber \\
 & =-E_{f,k_c}\left|0_f\right\rangle\left\langle 0_f|0_i\right\rangle+E_{f,k_c}\left|1_f\right\rangle\left\langle 1_f|0_i\right\rangle \nonumber\\
&  =-E_{f,k_c}\left|0_f\right\rangle\left\langle 0_f|0_i\right\rangle+E_{f,k_c}(1-\left|0_f\right\rangle\left\langle 0_f\right|)\left|0_i\right\rangle \nonumber\\
&  =-E_{f,k_c}\left|0_f\right\rangle\left\langle 0_f|0_i\right\rangle+E_{f,k_c}\left|0_i\right\rangle\nonumber\\
&\ \ \ \  -E_{f,k_c}\left|0_f\right\rangle\left\langle 0_f|0_i\right\rangle\nonumber\\
 &= -2E_{f,k_c}\left|0_f\right\rangle\left\langle 0_f|0_i\right\rangle+E_{f,k_c}\left|0_i\right\rangle\nonumber \\
 &= -2E_{f,k_c}(\left|0_i\right\rangle\left\langle 0_i\right|+\left|1_i\right\rangle\left\langle 1_i\right|)\left|0_f\right\rangle\left\langle 0_f|0_i\right\rangle \nonumber\\
&\ \ \ \  +E_{f,k_c}\left|0_i\right\rangle\nonumber\\
 & = -2E_{f,k_c}\left|0_i\right\rangle\left\langle 0_i|0_f\right\rangle\left\langle 0_f|0_i\right\rangle\nonumber\\
 & \ \ \ \ -2E_{f,k_c}\left|1_i\right\rangle\left\langle 1_i|0_f\right\rangle\left\langle 0_f|0_i\right\rangle+E_{f,k_c}\left|0_i\right\rangle\nonumber\\
 & =-2E_{f,k_c}\left|0_i\right\rangle\left|\left\langle 0_i|0_f\right\rangle\right|^2-e^{i\theta}E_{f,k_c}\left|1_i\right\rangle  \nonumber\\
& \ \ \ \ +E_{f,k_c}\left|0_i\right\rangle\nonumber\\
 & =-E_{f,k_c}\left|0_i\right\rangle-e^{i\theta}E_{f,k_c}\left|1_i\right\rangle+E_{f,k_c}\left|0_i\right\rangle \nonumber \\
 & =-e^{i\theta}E_{f,k_c}\left|1_i\right\rangle.\label{eq:Hfk0i}
\end{align}
As a result, we get $H_{f,k_{c}}\left|0_i\right\rangle \propto\left|1_i\right\rangle$. And it is easy to get $\left\langle 0_i\right|H_{f,k_{c}}\left|0_i\right\rangle=-e^{i\theta}E_{f,k_c}\left\langle 0_i|1_i\right\rangle=0$. In addition, we can derive
\begin{align}
\left\langle 0_i\right|H^2_{f,k_{c}}\left|0_i\right\rangle
& =\left\langle 0_i\right|H^2_{f,k_{c}}(\left|0_f\right\rangle\left\langle 0_f\right|+\left|1_f\right\rangle\left\langle 1_f\right|)\left|0_i\right\rangle  \nonumber\\
& =\left\langle 0_i\right|H^2_{f,k_{c}}\left|0_f\right\rangle\left\langle 0_f|0_i\right\rangle  \nonumber\\
& \ \ \ \ +\left\langle 0_i\right|H^2_{f,k_{c}}\left|1_f\right\rangle\left\langle 1_f|0_i\right\rangle \nonumber\\
 & =E^2_{f,k_c}\left\langle 0_i|0_f\right\rangle\left\langle 0_f|0_i\right\rangle \nonumber\\
& \ \ \ \ +E^2_{f,k_c}\left\langle 0_i|1_f\right\rangle\left\langle 1_f|0_i\right\rangle\nonumber\\
 & =E^2_{f,k_c}\left|\left\langle 0_i|0_f\right\rangle\right|^2+E^2_{f,k_c}\left|\left\langle 0_i|1_f\right\rangle\right|^2\nonumber\\
 & =E^2_{f,k_c}.
\end{align}

In the same way, we can derive
\begin{align}
H_{f,k_{c}}\left|1_i\right\rangle  & =H_{f,k_{c}}(\left|0_f\right\rangle\left\langle 0_f\right|+\left|1_f\right\rangle\left\langle 1_f\right|)\left|1_i\right\rangle \nonumber\\
& =H_{f,k_{c}}\left|0_f\right\rangle\left\langle 0_f|1_i\right\rangle+H_{f,k_{c}}\left|1_f\right\rangle\left\langle 1_f|1_i\right\rangle\nonumber \\
 & =-E_{f,k_c}\left|0_f\right\rangle\left\langle 0_f|1_i\right\rangle+E_{f,k_c}\left|1_f\right\rangle\left\langle 1_f|1_i\right\rangle \nonumber\\
&  =-E_{f,k_c}\left|0_f\right\rangle\left\langle 0_f|1_i\right\rangle+E_{f,k_c}(1-\left|0_f\right\rangle\left\langle 0_f\right|)\left|1_i\right\rangle \nonumber\\
&  =-E_{f,k_c}\left|0_f\right\rangle\left\langle 0_f|1_i\right\rangle+E_{f,k_c}\left|1_i\right\rangle\nonumber\\
&\ \ \ \  -E_{f,k_c}\left|0_f\right\rangle\left\langle 0_f|1_i\right\rangle\nonumber\\
 &= -2E_{f,k_c}\left|0_f\right\rangle\left\langle 0_f|1_i\right\rangle+E_{f,k_c}\left|1_i\right\rangle\nonumber \\
 &= -2E_{f,k_c}(\left|0_i\right\rangle\left\langle 0_i\right|+\left|1_i\right\rangle\left\langle 1_i\right|)\left|0_f\right\rangle\left\langle 0_f|1_i\right\rangle \nonumber\\
&\ \ \ \  +E_{f,k_c}\left|1_i\right\rangle\nonumber\\
 & = -2E_{f,k_c}\left|0_i\right\rangle\left\langle 0_i|0_f\right\rangle\left\langle 0_f|1_i\right\rangle\nonumber\\
 & \ \ \ \ -2E_{f,k_c}\left|1_i\right\rangle\left\langle 1_i|0_f\right\rangle\left\langle 0_f|1_i\right\rangle+E_{f,k_c}\left|1_i\right\rangle\nonumber\\
& = -2E_{f,k_c}\left|0_i\right\rangle(\left\langle 1_i|0_f\right\rangle\left\langle 0_f|0_i\right\rangle)^*\nonumber\\
 & \ \ \ \ -2E_{f,k_c}\left|1_i\right\rangle\left|\left\langle 1_i|0_f\right\rangle\right|^2+E_{f,k_c}\left|1_i\right\rangle\nonumber\\
 & = -e^{-i\theta}E_{f,k_c}\left|0_i\right\rangle\nonumber -E_{f,k_c}\left|1_i\right\rangle+E_{f,k_c}\left|1_i\right\rangle\nonumber\\
 & =-e^{-i\theta}E_{f,k_c}\left|0_i\right\rangle.\label{eq:Hfk1i}
\end{align}

If $d_{0,k_{c}}(\gamma_{i})$ is also equal to 0, then the eigenvalue equations of $H_{i,k_{c}}$ are
\begin{align}
H_{i,k_{c}}\left|0_f\right\rangle & =-E_{i,k_c}\left|0_i\right\rangle,\\
H_{i,k_{c}}\left|1_f\right\rangle & =E_{i,k_c}\left|1_i\right\rangle.
\end{align}
Here $E_{i,k_c}=\left|\mathbf{d}_{k_{c}}(\gamma_{i})\right|$. According to Eq. (\ref{eq:Hfk0i}) and Eq. (\ref{eq:Hfk1i}), we can get
\begin{align}
H_{i,k_{c}}H_{f,k_{c}}\left|0_i\right\rangle & =-e^{i\theta}E_{f,k_c}H_{i,k_{c}}\left|1_i\right\rangle \nonumber\\
& =-e^{i\theta}E_{f,k_c}E_{i,k_{c}}\left|1_i\right\rangle\nonumber \\
& =E_{i,k_{c}}(-e^{i\theta}E_{f,k_c}\left|1_i\right\rangle) \nonumber\\
& =E_{i,k_{c}}H_{f,k_{c}}\left|0_i\right\rangle \nonumber\\
& =H_{f,k_{c}}E_{i,k_{c}}\left|0_i\right\rangle\nonumber \\
& =-H_{f,k_{c}}H_{i,k_{c}}\left|0_i\right\rangle,
\end{align}
and
\begin{align}
H_{i,k_{c}}H_{f,k_{c}}\left|1_i\right\rangle & =-e^{-i\theta}E_{f,k_c}H_{i,k_{c}}\left|0_i\right\rangle \nonumber\\
& =e^{-i\theta}E_{f,k_c}E_{i,k_{c}}\left|0_i\right\rangle\nonumber \\
& =-E_{i,k_{c}}(-e^{-i\theta}E_{f,k_c}\left|0_i\right\rangle) \nonumber\\
& =-E_{i,k_{c}}H_{f,k_{c}}\left|1_i\right\rangle \nonumber\\
& =-H_{f,k_{c}}E_{i,k_{c}}\left|1_i\right\rangle\nonumber \\
& =-H_{f,k_{c}}H_{i,k_{c}}\left|1_i\right\rangle.
\end{align}
Because $\{\left|0_i\right\rangle,\left|1_i\right\rangle\}$ is a set of complete bases, we can draw a conclusion that
\begin{equation}
\{H_{i,k_{c}},H_{f,k_{c}}\}=0.
\end{equation}


\begin{thebibliography}{10}
\bibitem{Sachdev} S. Sachdev, Quantum Phase Transitions (Cambridge University Press, Cambridge, 1999).

\bibitem{Zanardi2006} P. Zanardi and N. Paunkovic, Ground state overlap and quantum phase transitions, Phys. Rev. E \textbf{74}, 031123 (2006).

\bibitem{You2007} W. L. You, Y. W. Li, and S. J. Gu, Fidelity, dynamic structure factor, and susceptibility in critical phenomena, Phys. Rev. E
\textbf{76}, 022101 (2007).

\bibitem{Zanardi2007} P. Zanardi, M. Cozzini, and P. Giorda, Ground state fidelity and quantum phase transitions in free Fermi systems, J. Stat. Mech.
(2007) L02002.

\bibitem{Buonsante2007} P. Buonsante and A. Vezzani, Ground-state fidelity and bipartite entanglement in the Bose-Hubbard model, Phys. Rev. Lett. \textbf{98}, 110601 (2007).

\bibitem{Cozzini2007-1} M. Cozzini, R. Ionicioiu, and P. Zanardi, Quantum fidelity and quantum phase transitions in matrix product states, Phys. Rev. B \textbf{76}, 104420 (2007)

\bibitem{Cozzini2007-2} M. Cozzini, P. Giorda, and P. Zanardi, Quantum phase transitions and quantum fidelity in free fermion graphs, Phys. Rev. B \textbf{75}, 014439 (2007)

\bibitem{YanfMF2007} M.-F. Yang, Ground-state fidelity in one-dimensional gapless models, Phys. Rev. B \textbf{76}, 180403(R) (2007).

\bibitem{ZhouHQ2008} H.-Q. Zhou, J.-H. Zhao, and B. Li, Fidelity approach to quanum phase transitions: Finite-size scaling for the quantum Ising model in a transverse field, J. Phys. A \textbf{41}, 492002 (2008).

\bibitem{ChenS2008} S. Chen, L. Wang, Y. Hao, and Y. Wang, Intrinsic relation between ground-state fidelity and the characterization of a quantum phase transition, Phys. Rev. A \textbf{77}, 032111 (2008)

\bibitem{ZRams2011} M. M. Rams and B. Damski, Quantum fidelity in the thermodynamic limit, Phys. Rev. Lett. \textbf{106}, 055701 (2011).

\bibitem{Dutta} V. Mukherjee, A. Polkovnikov, and A. Dutta, Oscillating fidelity susceptibility near a quantum multicritical point,
Phys. Rev. B \textbf{83}, 075118 (2011).

\bibitem{SunG2015} G. Sun, A. K. Kolezhuk, and T. Vekua, Fidelity
at Berezinskii-Kosterlitz-Thouless quantum phase transitions, Phys.
Rev. B \textbf{91}, 014418 (2015).

\bibitem{SunG} G. Sun, B.-B. Wei, and S.-P. Kou, Fidelity as a probe
for a deconfined quantum critical point, Phys. Rev. B \textbf{100},
064427 (2019).

\bibitem{Zeng2024} Y. Zeng, B. Zhou, and S. Chen, Exact zeros of fidelity in finite-size systems as a signature for probing quantum phase transitions, Phys. Rev. E \textbf{109}, 064130 (2024).

\bibitem{Anderson1967} P. W. Anderson, Infrared Catastrophe in Fermi Gases with Local Scattering Potentials, Phys. Rev. Lett.  \textbf{18}, 1049 (1967).

\bibitem{Heyl2013PRL} M. Heyl, A. Polkovnikov and S. Kehrein, dynamical quantum phase transitions in the transverse-field Ising model, Phys. Rev. Lett. \textbf{110}, 135704 (2013).

\bibitem{Karrasch2013PRB} C. Karrasch and D. Schuricht, Dynamical phase transitions after quenches in nonintegrable models, Phys. Rev. B \textbf{87}, 195104 (2013).

\bibitem{Kriel2014PRB}  J. N. Kriel, C. Karrasch, and S. Kehrein, Dynamical quantum phase transitions in the axial next-nearest-neighbor Ising chain, Phys. Rev. B \textbf{90}, 125106 (2014).

\bibitem{Heyl2015RPL} M. Heyl, Scaling and Universality at Dynamical Quantum Phase Transitions, Phys. Rev. Lett. \textbf{115}, 140602 (2015).


\bibitem{Dora2015PRB} S. Vajna and B. D\'{o}ra, Topological classification of dynamical phase transitions, Phys. Rev. B \textbf{91}, 155127 (2015).

\bibitem{Schmitt2015PRB} M. Schmitt and S. Kehrein, Dynamical quantum phase transitions in the Kitaev honeycomb model, Phys. Rev. B \textbf{92}, 075114 (2015).

\bibitem{HuangZ2016PRB} Z. Huang and A. V. Balatsky, Dynamical quantum phase transitions: Role of topological nodes in wave function overlaps, Phys. Rev. Lett. \textbf{117}, 086802 (2016).

\bibitem{Zvyagin2016LTP}  A. A. Zvyagin, Dynamical quantum phase transitions (Review Article), Low Temp. Phys. \textbf{42}, 971-994 (2016).

\bibitem{Bhattacharya2017PRB} U. Bhattacharya and A. Dutta, Emergent topology and dynamical quantum phase transitions in two-dimensional closed quantum systems, Phys. Rev. B \textbf{96}, 014302 (2017).

\bibitem{Bhattacharjee2018PRB} S. Bhattacharjee and A. Dutta, Dynamical quantum phase transitions in extended transverse Ising models, Phys. Rev. B \textbf{97}, 134306 (2018).

\bibitem{Heyl2018RPP} M. Heyl, Dynamical quantum phase transitions: a review, Rep. Prog. Phys. \textbf{81}, 054001 (2018).

\bibitem{ZhouBZ2019} B. Zhou, C. Yang, and S. Chen, Signature of a nonequilibrium quantum phase transition in the long-time
average of the Loschmidt echo, Phys. Rev. B \textbf{100},
184313 (2019).

\bibitem{Corps} A. L. Corps and A. Relano, Theory of dynamical phase transitions in quantum systems with symmetry-breaking eigenstates,
Phys. Rev. Lett. \textbf{130}, 100402 (2023).

\bibitem{Cheraghi2023NJP} H. Cheraghi and N. Sedlmayr, Dynamical quantum phase transitions following double quenches: persistence of the initial state vs dynamical phases, New J. Phys. \textbf{25}, 103035 (2023).

\bibitem{ZhouBZ2021} B. Zhou, Y. Zeng, and S. Chen, Exact zeros of
the Loschmidt echo and quantum speed limit time for the dynamical
quantum phase transition in finite-size systems, Phys. Rev. B \textbf{104}, 094311 (2021).

\bibitem{ZengYM} Y. Zeng, B. Zhou, and S. Chen, Dynamical singularity of the rate function for quench dynamics in finite-size quantum systems, Phys. Rev. B \textbf{107}, 134302 (2023).

\bibitem{YangC} C. Yang, L. Li and S. Chen, Dynamical topological
invariant after a quantum quench, Phys. Rev. B \textbf{97}, 060304(R) (2018).


\bibitem{Flaschner} N. Fl\"{a}schner, D. Vogel, M. Tarnowski, et al. Observation of dynamical vortices after quenches in a system with topology. Nat. Phys. \textbf{14}, 265-268 (2018).


\bibitem{Liska} D. Liska and V. Gritsev, The Loschmidt Index, SciPost Phys. \textbf{10}, 100 (2021).

\bibitem{WongCY2024PRB} C. Y. Wong, T. H. Hui, P. D. Sacramento, and W. C. Yu, Entanglement in quenched extended Su-Schrieffer-Heeger model with anomalous dynamical quantum phase transitions, Phys. Rev. B \textbf{110}, 054312 (2024).

\bibitem{Homrighausen2017} I. Homrighausen, N. O. Abeling, V. Zauner-Stauber, and J. C. Halimeh, Anomalous dynamical phase in quantum spin chains with long-range interactions, Phys. Rev. B \textbf{96}, 104436 (2017).

\bibitem{Halimeh2020}J. C. Halimeh, M. V. Damme, V. Zauner-Stauber, and L. Vanderstraeten, Quasiparticle origin of dynamical quantum phase transitions, Phys. Rev. Research \textbf{2}, 033111 (2020).

\bibitem{Vajna2014PRB} S. Vajna and B. D\'{o}ra, Disentangling dynamical phase transitions from equilibrium phase transitions, Phys. Rev. B \textbf{89}, 161105(R) (2014).

\bibitem{Andraschko2014PRB} F. Andraschko and J. Sirker, Dynamical quantum phase transitions and the Loschmidt echo: A transfer matrix approach, Phys. Rev. B \textbf{89}, 125120 (2014).

\bibitem{Sharma2015PRB}  S. Sharma, S. Suzuki, and A. Dutta, Quenches and dynamical phase transitions in a nonintegrable quantum Ising model, Phys. Rev. B \textbf{92}, 104306 (2015).

\bibitem{Jafari2019SR} R. Jafari, Dynamical Quantum Phase Transition and Quasi Particle Excitation, Sci Rep  \textbf{9}, 2871 (2019).

\bibitem{Sadrzadeh2021PRB} M. Sadrzadeh, R. Jafari, and A. Langari, Dynamical topological quantum phase transitions at criticality, Phys. Rev. B \textbf{103}, 144305 (2021).

\bibitem{Stumper2022PRR} S. Stumper, M. Thoss, and J. Okamoto, Interaction-driven dynamical quantum phase transitions in a strongly correlated bosonic system, Phys. Rev. Research \textbf{4}, 013002 (2022).

\bibitem{Rossi2022PRB} L. Rossi and F. Dolcini, Nonlinear current and dynamical quantum phase transitions in the flux-quenched Su-Schrieffer-Heeger model, Phys. Rev. B \textbf{106}, 045410 (2022).

\bibitem{Divakaran2016PRE} U. Divakaran, S. Sharma, and A. Dutta, Tuning the presence of dynamical phase transitions in a generalized \textit{XY} spin chain, Phys. Rev. E \textbf{93}, 052133 (2016).

\bibitem{Halimeh2017PRB} J. C. Halimeh and V. Zauner-Stauber, Dynamical phase diagram of quantum spin chains with long-range interactions, Phys. Rev. B \textbf{96}, 134427 (2017).

\bibitem{Zauner-Stauber2017PRE} V. Zauner-Stauber and J. C. Halimeh, Probing the anomalous dynamical phase in long-range quantum spin chains through Fisher-zero lines, Phys. Rev. E \textbf{96}, 062118 (2017).

\bibitem{Obuchi2017PRB} T. Obuchi, S. Suzuki, and K. Takahashi, Complex semiclassical analysis of the Loschmidt amplitude and dynamical quantum phase transitions, Phys. Rev. B \textbf{95}, 174305 (2017).

\bibitem{Lang2018PRB} J. Lang, B. Frank, and J. C. Halimeh, Concurrence of dynamical phase transitions at finite temperature in the fully connected transverse-field Ising model, Phys. Rev. B \textbf{97}, 174401 (2018).

\bibitem{Uhrich2020PRB} P. Uhrich, N. Defenu, R. Jafari, and J. C. Halimeh, Out-of-equilibrium phase diagram of long-range superconductors, Phys. Rev. B \textbf{101}, 245148 (2020).

\bibitem{Lieb1961AP} E. Lieb, T. Schultz, and D. Mattis, Two soluble models of an antiferromagnetic chain, Ann. Phys. (NY) \textbf{16}, 407 (1961).

\bibitem{Bunder1999PRB} J. E. Bunder and R. H. McKenzie, Effect of disorder on quantum phase transitions in anisotropic XY spin chains in a transverse field, Phys. Rev. B \textbf{60}, 344 (1999).

\bibitem{LuoQ2018PRE} Q. Luo, J. Zhao, and X. Wang, Fidelity susceptibility of the anisotropic \textit{XY} model: The exact solution, Phys. Rev. E \textbf{98}, 022106 (2018).



\end{thebibliography}
\end{document}